\documentclass{icrc}
\usepackage{amssymb,amsmath}
\usepackage{times}
\newif\ifpdf 
\ifx\pdfoutput\undefined
\pdffalse    
\else
\pdfoutput=1 
\pdftrue
\fi \ifpdf 
    \usepackage[pdftex]{graphicx} 
\else
    \usepackage{graphicx}         
\fi

\begin{document}

\title{Atmospheric neutrino flux and muon data}
\author[1]{G. Fiorentini}
\author[1,2]{V. A. Naumov}
\affil[1]{Dipartimento di Fisica and Sezione INFN di Ferrara,
          Via del Paradiso 12, I-44100 Ferrara, Italy}
\author[1]{F. L. Villante}
\affil[2]{Laboratory for Theoretical Physics, Irkutsk State
          University, Gagarin boulevard 20, RU-664003 Irkutsk,
          Russia}
\correspondence{V. A. Naumov (naumov@fe.infn.it)}
\runninghead{G.\,Fiorentini, V.\,A.\,Naumov, and F.\,L.\,Villante:
             Atmospheric neutrino flux and muon data}
\firstpage{1}
\pubyear{2001}
\maketitle

\begin{abstract}
We present a new one-dimensional calculation of low and intermediate
energy atmospheric muon and neutrino fluxes, using up-to-date data on
primary cosmic rays and hadronic interactions. The existing agreement
between calculated muon fluxes and the data of the CAPRICE\,94 muon
experiment provides an evidence in favor of the validity of our
description of hadronic interactions and shower development. This
also supports our neutrino fluxes which are essentially lower than
those used for the standard analyses of the sub-GeV and multi-GeV
neutrino induced events in underground detectors.
\end{abstract}

\protect\section{Introduction} \label{sec:Introduction}

In this work we present some results of a new calculation of low and
intermediate energy muon and neutrino fluxes based on up-to-date data
on primary cosmic-ray flux and hadronic interactions. In order to
demonstrate the validity of our approach, we give a comparison of the
predicted muon fluxes with the recent data by the CAPRICE\,94
balloon-borne experiment \citep{Boezio00}.

Our calculations are based on an updated code CORT
\citep{Fiorentini01}. Like the earlier version (see
\citet{Bugaev87,Bugaev89,Bugaev90} and references therein) the new
code implements a numerical solution of a system of one-dimensional
(1D) kinetic equations describing the propagation of cosmic-ray
nuclei, nucleons, $\pi$ and $K$ mesons, muons, neutrinos, and
antineutrinos of low and intermediate energies through a spherical,
nonisothermal atmosphere.

In order to evaluate geomagnetic effects and to take into account the
anisotropy of the primary cosmic-ray flux in the vicinity of the
Earth, we use the method by \citet{Naumov84} and detailed maps of the
effective vertical cutoff rigidities by \citet{Dorman71}. The maps
are corrected for the geomagnetic pole drift and compared with the
later results reviewed by \citet{Smart94} and with the recent AMS
data on the proton flux in near earth orbit \citep{Alcaraz00a}. The
interpolation between the reference points of the maps is performed
by means of two-dimensional local B-spline. The Quenby-Wenk relation
\citep{Dorman71}, re-normalized to the vertical cutoffs, is applied
for evaluating the effective cutoffs for oblique directions. More
sophisticated effects, like the short-period variations of the
geomagnetic field, Forbush decrease, re-entrant cosmic-ray albedo
contribution, etc., are neglected. We also neglect the geomagnetic
bending of the trajectories of charged secondaries and multiple
scattering effects. Validity of our treatment of propagation of
secondary nucleons and nuclei was confirmed using all available data
on the proton and neutron spectra in the atmosphere
\citep{Naumov84,Bugaev85}.

The meteorological effects are included using the Dorman model of the
atmosphere \citep{Dorman72} which assumes an isothermal stratosphere
and constant gradient of temperature (as a function of depths) below
the tropopause. Ionization, radiative and photonuclear muon energy
losses are treated as continuous processes. This approximation is
quite tolerable for atmospheric depths
$h\lesssim2\times10^3$ g/cm${}^2$ at all energies of interest
\citep{Naumov94}. Propagation of $\mu^+$ and $\mu^-$ originating from
every source (pion or kaon decay) is described by separate kinetic
equations for muons with definite polarization at production. These
equations automatically account for muon depolarization through the
energy loss (but not through the Coulomb scattering).

\protect\section{Primary cosmic ray spectrum and composition}
\label{sec:Primaries}

In the present calculations, the nuclear component of primary cosmic
rays is broken up into 5 principal groups: H, He, CNO, Ne-S and Fe
with average atomic masses $A$ of 1, 4, 15, 27 and 56, respectively.
We do not take into account the isotopic composition of the primary
nuclei and assume $Z=A/2$ for $A>1$, since the expected effect on the
secondary lepton fluxes is estimated to be small with respect to
present-day experimental uncertainties in the absolute cosmic-ray
flux and chemical composition.

We parametrize the spectra of the H and He groups at
$E<120$ GeV/nucleon by fitting the data of the balloon-borne
experiment BESS obtained by a flight in 1998 \citep{Sanuki00}. For
higher energies (but below the knee) we use data by a series of
twelve balloon flights of JACEE \citep{Asakimori98} and the result of
an analysis by \citet{Wiebel-Sooth98} based upon a representative
compilation of world data on primaries. We assume that the spectra of
the remaining three nuclear groups are similar to the helium
spectrum. This assumption does not contradict the world data for the
CNO and Ne-S nuclear groups but works a bit worse for the iron group.
Nevertheless, a more sophisticated model would be unpractical since
the corresponding correction would affect the secondary lepton fluxes
by a negligible margin.

In this paper we do not consider the effects of solar modulation.
Therefore the predicted muon and neutrino fluxes are to some extent
the maximum ones possible within our approach.

\protect\section{Nucleon-nucleus and nucleus-nucleus interactions}
\label{sec:interactions}

All calculations with the earlier version of CORT (see, e.g.,
\citet{Naumov84,Bugaev85,Bugaev87,Bugaev89,Bugaev90,Bugaev98}) were
based on semiempirical models for inclusive nucleon and meson
production in collisions of nucleons with nuclei by
\citet{Kimel'74,Kimel'75,Serov73}.%
\footnote{See also \citet{Kalinovsky85,Sychev99} for the most
          recent versions.}
The Kimel'-Mokhov (KM) model is valid for projectile nucleon momenta
above $\sim4$ GeV/c and for the secondary nucleon, pion and kaon
momenta above 450, 150 and 300 MeV/c, respectively. Outside these
ranges (that is mainly within the region of resonance production of
pions) the Serov-Sychev (SS) model was used.

Both models are in essence comprehensive parametrizations of the
relevant accelerator data. In our opinion, the combined ``KM+SS''
model provides a rather safe and model-independent basis to the
low-energy atmospheric muon and neutrino calculations. However it is
not free of uncertainties.  For the present calculation, the fitting
parameters of the KM model for meson and nucleon production off
different nuclear targets were updated using accelerator data not
available for the original analysis
\citep{Kimel'74,Kimel'75,Kalinovsky85}. The values of the parameters
were extrapolated to the air nuclei (N, O, Ar, C).  The overall
correction is less than 10-15\% within the kinematic regions
significant to atmospheric cascade calculations. Besides that
energy-dependent correction factors were introduced into the model to
tune up the output $\pi^+/\pi^-$ ratio taking into account the
relevant new data.

The processes of meson regeneration and charge exchange
($\pi^\pm+\mathrm{Air}\to\pi^{\pm(\mp)}+X$ etc.) are not of critical
importance for production of leptons with energies of our interest
and can be considered in a simplified way. Here we use a proper
renormalization of the meson interaction lengths, which was deduced
from the results by \citet{Vall86} obtained for high-energy cascades.

The next important ingredient of any cascade calculations is a model
for nucleus-nucleus collisions. Here we consider a modest
generalization of a simple ``Glauber-like'' model used in
\citep{Naumov84,Bugaev85}. Namely, we write the inclusive spectrum of
secondary particles $c$ ($c=p$, $n$, $\pi^\pm$, $K^\pm$,
$K^0,\ldots$) produced in AB collisions as
\begin{align*}
\frac{dN_{\mathrm{AB} \to cX}}{d x} = & \xi_{\mathrm{AB}}^c
\left[Z \frac{dN_{p\mathrm{B} \to cX}}{dx}+
   (A-Z)\frac{d N_{n\mathrm{B} \to cX}}{dx}\right]\nonumber\\
&+\left(1-\xi_{\mathrm{AB}}^c\right)
\left[Z\delta_{cp}+(A-Z)\delta_{cn}\right]\delta(1-x).
\end{align*}
Here $dN_{N\mathrm{B} \to cX}/dx$ is the spectrum of particles $c$
produced in $N\mathrm{B}$ collisions ($N=p,n$) and
$\xi_{\mathrm{AB}}^c$ is the average fraction of inelastically
interacting nucleons of the projectile nucleus A. The term
proportional to delta function describes the contribution of
``spectator'' nucleons from the projectile nucleus.

In the standard Glauber-Gribov multiple scattering theory the
quantity $\xi_{\mathrm{AB}}^c$ is certainly independent of the type
of inclusive particle $c$. On the other hand, it depends of the type
of nucleus collision. Indeed, essentially all nucleons participate in
the central AB collisions ($\xi_{\mathrm{AB}}^c\simeq1$)%
\footnote{Here we suppose for simplicity that the atomic weight of
          the projectile nucleus is not much larger than that of
          the target nucleus.}
while, according to the well-known Bialas--Bleszy\'nski--Czy\.z (BBC)
relation \citep{Bialas76},
$\xi_{\mathrm{AB}}^c=\sigma_{N\mathrm{B}}^{\mathrm{inel}}/
                     \sigma_{\mathrm{AB}}^{\mathrm{inel}}$
for the minimum bias collisions.

To use the above model in a cascade calculation one should take into
account that nucleons and mesons are effectively produced in nuclear
collisions of different kind. Namely, the contribution from central
collisions is almost inessential for the nucleon component of the
cascade but quite important for light meson production. Thus one can
expect that effectively
$\xi_{\mathrm{AB}}^{\pi,K}>\xi_{\mathrm{AB}}^{p,n}$. We use the BBC
relation for nucleon production by any nucleus while for meson
production we put $\xi_{\mathrm{He-Air}}^{\pi,K}=\xi$, where $\xi$ is
a free parameter. Variations of this parameter within the
experimental limits yield a comparatively small effect to the muon
fluxes (except for very high altitudes) and inessential
($\lesssim6$\%) effect to the neutrino fluxes at sea level
\citep{Fiorentini01}. Effect of similar variations of the parameters
$\xi_{\mathrm{A-Air}}^{\pi,K}$ for other nuclear groups is completely
negligible. Below, we use the fixed value $\xi=0.685$.

\protect\section{Numerical results and discussion} \label{sec:NumRes}

\begin{figure}[tb]
\vspace*{2.0mm} 
\ifpdf\includegraphics[width=8.3cm]{c94f.pdf} 
\else \includegraphics[width=8.3cm]{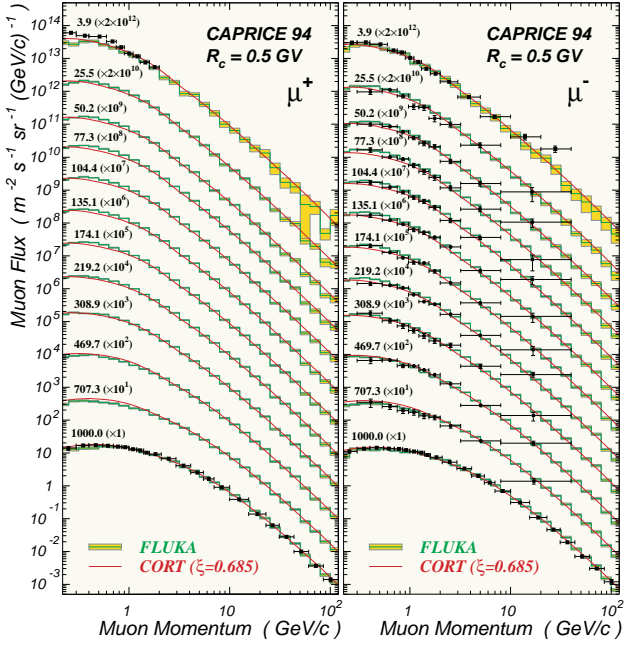} 
\fi
\protect\caption{Differential momentum spectra of $\mu^+$ and $\mu^-$
                 for 12 atmospheric depths. The data points are from
                 CAPRICE\,94 experiment \citep{Boezio00}. The curves
                 and histograms are calculations with CORT and FLUKA,
                 respectively, performed for the conditions of the
                 experiment. The numbers indicate the FAD
                 (in g/cm$^2$) and scale factors (in parentheses).
\label{f:CAPRICE94a}}
\end{figure}

\begin{figure}[tb]
\vspace*{2.0mm} 
\ifpdf\includegraphics[width=8.3cm]{c94d.pdf} 
\else \includegraphics[width=8.3cm]{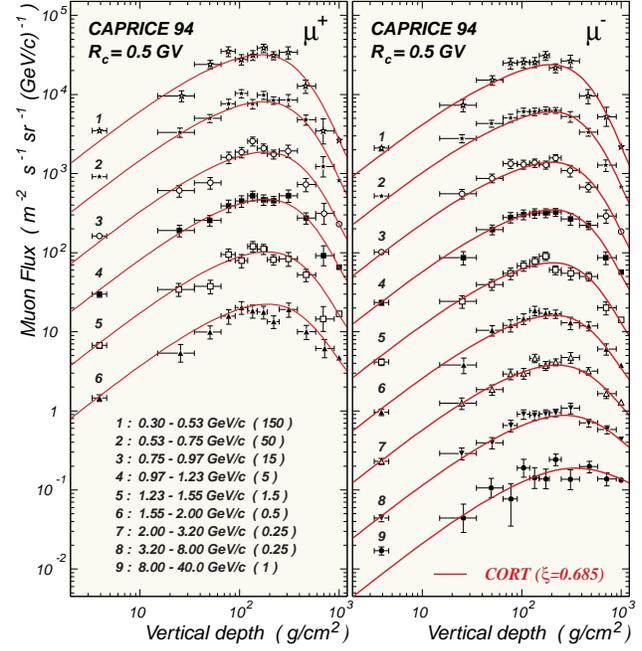} 
\fi
\protect\caption{Atmospheric growth curves for $\mu^+$ and $\mu^-$.
                 The data points are from the CAPRICE\,94 experiment
                 \citep{Boezio00}. The curves are calculations with
                 CORT performed for the conditions of the experiment.
                 The legend shows the muon momentum bins and scale
                 factors (in parentheses) for both panels.
\label{f:CAPRICE94b}}
\end{figure}

In Fig. \ref{f:CAPRICE94a} we compare the calculated momentum spectra
of $\mu^+$ and $\mu^-$ for twelve atmospheric depths with the data of
the balloon-borne experiment CAPRICE\,94 \citep{Boezio00}.  The
values of depths indicated in Fig. \ref{f:CAPRICE94a} are the
Flux-weighted Average Depths (FAD) \citep{Boezio00} for 11
atmospheric ranges $\Delta h_i$ ($h_i<10^3$ g/cm$^2$).%
\footnote{We neglect the small spreads of the FADs within some of the
          ranges $\Delta h_i$.}
Figure \ref{f:CAPRICE94a} also includes the result of the recent 3D
calculation by \citet{Battistoni01} based on the FLUKA 3D Monte Carlo
code \citep{Battistoni00}. Figure \ref{f:CAPRICE94b} shows the
atmospheric growth of muon fluxes for six momentum bins for $\mu^+$
and nine bins for $\mu^-$. The CAPRICE\,94 data \citep{Boezio00} are
compared with our calculations performed for the same bins.

From Figs. \ref{f:CAPRICE94a} and \ref{f:CAPRICE94b}, one concludes
that there is a substantial agreement between the CORT predictions
and the current muon data within wide ranges of muon momenta and
atmospheric depths. In particular, the agreement is good for the
region of effective production of leptons (100--300 g/cm${}^2$), in
which the spread of the data is minimal. This provides an evidence
for the validity of our nuclear-cascade model. The comparison
between CORT and FLUKA presented in Fig. \ref{f:CAPRICE94a}
demonstrates a very good agreement (see also the result by
\citet{Poirier01} obtained with FLUKA).

\begin{figure}[hb]
\vspace*{2.0mm} 
\ifpdf\includegraphics[width=8.3cm]{tab.pdf} 
\else \includegraphics[width=8.3cm]{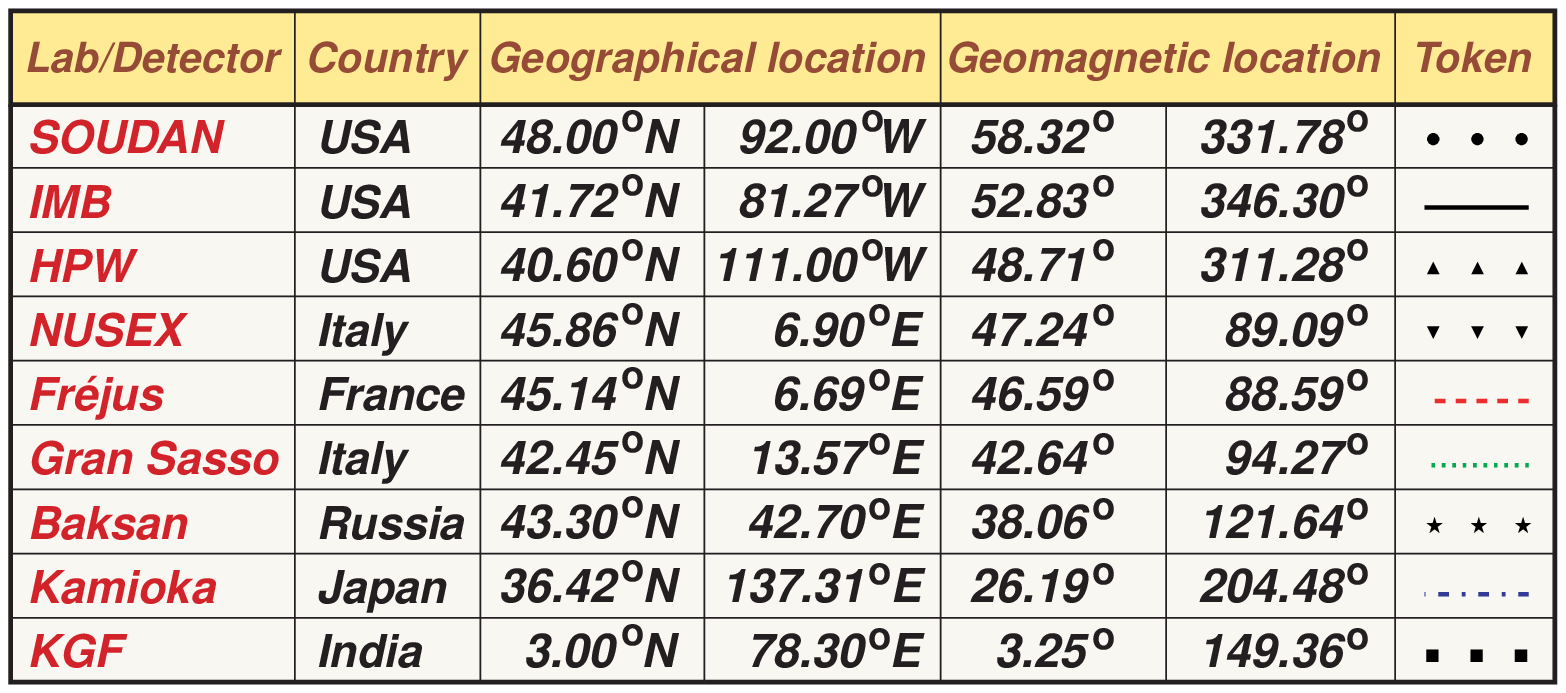} 
\fi
\protect\caption{List of some underground laboratories.
\label{f:Table}}
\end{figure}

\begin{figure}[t]
\vspace*{2.0mm} 
\ifpdf\includegraphics[width=8.3cm]{a.pdf} 
\else \includegraphics[width=8.3cm]{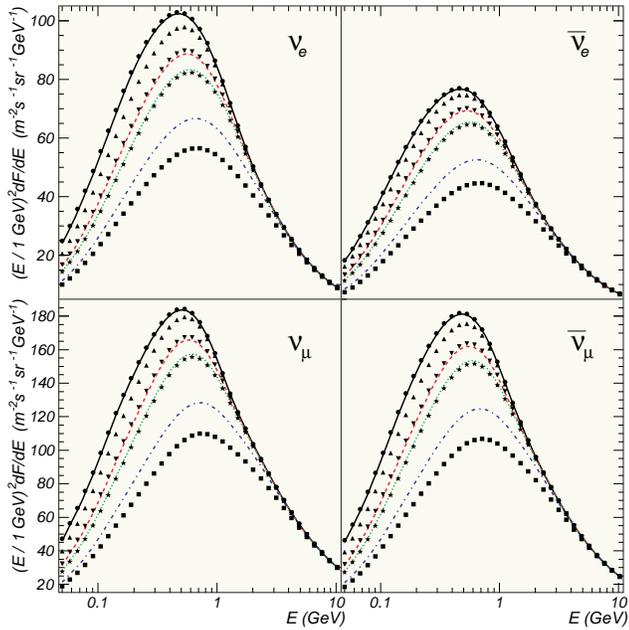} 
\fi
\protect\caption{Scaled $4\pi$ averaged fluxes of $\nu_e$,
                 $\overline{\nu}_e$, $\nu_\mu$, and
                 $\overline{\nu}_\mu$ for nine underground
                 laboratories (see Fig. \ref{f:Table} for the
                 notation).
\label{f:AverageFluxes}}
\end{figure}

\begin{figure}[t]
\vspace*{2.0mm} 
\ifpdf\includegraphics[width=8.3cm]{r.pdf} 
\else \includegraphics[width=8.3cm]{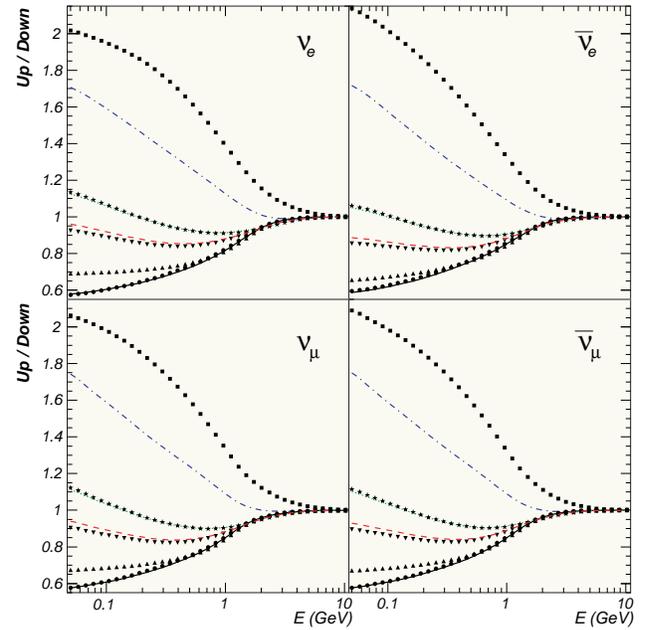} 
\fi
\protect\caption{Up-to-Down ratios of the $\nu_e$,
                 $\overline{\nu}_e$, $\nu_\mu$, and
                 $\overline{\nu}_\mu$ fluxes for nine underground
                 laboratories (see Fig. \ref{f:Table} for the
                 notation).
\label{f:UDown}}
\end{figure}

Now, let us consider our results for atmospheric neutrino (AN) fluxes
for the nine underground laboratories listed in Fig. \ref{f:Table}.
Figure \ref{f:AverageFluxes} shows the $\nu_e$, $\overline{\nu}_e$,
$\nu_\mu$ and $\overline{\nu}_\mu$ energy spectra averaged over all
zenith and azimuth angles. The ratios of the AN fluxes averaged over
the lower and upper semispheres (``up-to-down'' ratios) are shown in
Fig. \ref{f:UDown}. As a result of geomagnetic effects, both the
spectra and the up-to-down ratios are different for different
underground neutrino experiments.



Below 1-2 GeV our calculations lead to AN fluxes which are
essentially lower than those obtained by \citet{Barr89} and by
\citet{Honda90} and those are used in many analyses of the sub-GeV
and multi-GeV $\nu$ induced events in underground detectors. On the
other hand, our fluxes are rather close to the results obtained with
the earlier version of CORT \citep{Bugaev87,Bugaev89,Bugaev90}.
The AN fluxes and angular distributions calculated with CORT and
FLUKA are in a good agreement above 0.7--0.8 GeV (the difference is
typically $\lesssim15$\%) and disagreement (up to 30--40\%) at lower
energies.
A comparison between the earlier calculations of the AN fluxe is
discussed by \citet{Gaisser96}.

Here we do not discuss the complex problem of the interpretation of
the AN anomaly%
\footnote{For a recent review see Ref. \citep{Kajita01}.}
in the light of our result. However, we remark that
our low AN flux, when it is applied to the analysis of the
underground neutrino data, results in some electron excess together
with (or rather than) the muon deficit in the $\nu$ induced events.
We emphasize that in the context of our model it is difficult to
increase the AN flux without spoiling the agreement with the current
data on muon fluxes, hadronic cross sections, and primary cosmic-ray
spectrum.

\begin{acknowledgements}
We thank G. Battistoni, M. Boezio, and M. Circella for providing us
with their results in advance of publication.
V.\,A.\,N. is supported in part by the Ministry of Education of the
Russian Federation under grant No.015.02.01.004 (the program
``Universities of Russia -- Basic Researches'').
\end{acknowledgements}

\end{document}